\documentstyle[12pt, epsf]{article}
\renewcommand{\baselinestretch}{1.4}
\newcommand{\rhohat}{{\hat{\rho}}}
\newcommand{\bt}{{\beta}}

\newcommand{\bp} {{\bf p}}

\newcommand{\br} {{\bf r}}
\newcommand{\ggg} {{g_{\rho\pi\pi}}}
\newcommand{\fs} {{g_{b_1\rho\eta}^S}}
\newcommand{\fd} {{g_{b_1\rho\eta}^D}}
\newcommand{\fsw} {{g_{b_1\omega\pi}^S}}
\newcommand{\fdw} {{g_{b_1\omega\pi}^D}}
\newcommand{\hs} {{g_{\rhohat b_1\pi}^S}}
\newcommand{\hd} {{g_{\rhohat b_1\pi}^D}}
\newcommand{\cc}{\varsigma_1}
\newcommand{\dd}{\varsigma_2}
\newcommand{\ee}{\varsigma_3}
\newcommand{\qa}{\varpi_1}
\newcommand{\qd}{\varpi_2}
\newcommand{\qe}{\varpi_3}
\newcommand{\cm}{{\cal M}}

\newcommand{\btab}{\begin{tabbing}}
\newcommand{\etab}{\end{tabbing}}

\newcommand{\beqn}{\begin{equation}}
\newcommand{\eeqn}{\end{equation}}
\newcommand{\barr}[1]{\begin{array}{#1}}
\newcommand{\earr}{\end{array}}
\newcommand{\beqna}{\begin{eqnarray}}
\newcommand{\eeqna}{\end{eqnarray}}
\newcommand{\btablec}{\begin{table} \begin{center}}
\newcommand{\etablec}{\end{center} \end{table}}

\newcommand{\gapproxeq}{\lower.7ex\hbox{$\;\stackrel{\textstyle>}
{\sim}\;$}}
\newcommand{\plabel}[1]{\label{#1}}
\newcommand{\pbibitem}[1]{\bibitem{#1}}
\input epsf

\begin{document}
\title{
\begin{flushright} 
\small{hep-ph/9808225}\\
\small{JLAB-THY-98-20} \end{flushright} 
\vspace{0.6cm}  
\Large\bf Interpretation of Experimental $J^{PC}$ Exotic Signals}
\vskip 0.2 in
\author{Alexander Donnachie\thanks{\small \em E-mail: 
ad@a3.ph.man.ac.uk. Current address: Department of Physics and 
Astronomy, University of Manchester, Manchester M13 9PL, UK. } 
\hspace{.1cm}   and   Philip R. Page\thanks{\small \em E-mail:
prp@t5.lanl.gov. Current address: T-5, MS-B283, Los Alamos
National Laboratory, P.O. Box 1663, Los Alamos, NM 87545, USA.} \\
{\small \em Theory Group, Thomas Jefferson National Accelerator 
Facility,}\\ 
{\small \em 12000 Jefferson Avenue, Newport News, VA 23606, USA}}
\date{June 1998}
\maketitle
\begin{abstract}{We investigate theoretical interpretations of the 
1.4 GeV $J^{PC}$ exotic resonance reported by the E852 collaboration.
It is argued that interpretation in terms of a hybrid meson is 
untenable. A K--matrix analysis shows that the 1.4 GeV enhancement 
in the E852 $\eta\pi$ data can be understood as an interference of 
a non--resonant Deck--type background and a resonance 
at 1.6 GeV. A final state rescattering calculation shows that 
the 1.6 GeV hybrid has a $\eta\pi$ width which is bounded above by
$57\pm 14$ MeV.}
\end{abstract}
\bigskip


PACS number(s):  \hspace{.2cm} 12.39.Mk \hspace{.2cm}12.40.Nn \hspace{.2cm}
13.25.Jx\hspace{.2cm}13.75.Gx\hspace{.2cm}14.40.Cs\hspace{.2cm} 

\section{Introduction}

Evidence for a $J^{PC}=1^{-+}$ isovector resonance $\rhohat(1405)$ at 1.4 
GeV in the reaction $\pi^{-} p \rightarrow \eta\pi^{-}p$ has
been published recently by the E852 collaboration at 
BNL~\cite{bnletapi}. The mass and width quoted are $1370 \pm 
16^{+50}_{-30}$ MeV 
and $385\pm 40 ^{+65}_{-105}$ respectively.
These conclusions are strengthened by the claim of the Crystal Barrel 
collaboration
that there is evidence for the same resonance in $p\bar{p}$ annihilation 
with a mass of
$1400\pm 20\pm 20$ MeV and a width of $310\pm 50 ^{+50}_{-30}$ MeV 
\cite{cbar}, consistent with E852. However, the Crystal Barrel state 
is not seen as a peak in the $\eta\pi$ mass distribution, but is deduced
from interference in the Dalitz plot. 
Since the $J^{PC}$ of this state is ``exotic'', i.e. it implies 
that it is {\it not}
a conventional meson, considerable excitement has been generated,
particularly because the properties of the state appear to be in conflict 
with
theoretical expectations. The resonance is reported in natural parity 
exchange in the E852 experiment, and no statement can currently
be made about its production in unnatural parity exchange.

In addition there are two independent indications of a more massive
isovector $J^{PC}=1^{-+}$ exotic resonance $\rhohat(1600)$ in $\pi^{-} N 
\rightarrow
\pi^{+}\pi^{-}\pi^{-} N$.
The E852 collaboration recently reported evidence for a resonance at
$1593\pm 8^{+29}_{-47}$ MeV with a width of $168\pm 20^{+150}_{-12}$
 MeV~\cite{bnl97}.
These parameters are consistent with
the preliminary claim by the VES collaboration of a resonance at 
$1.62 \pm 0.02$ GeV with a width of
$0.24\pm 0.05$ GeV~\cite{ves93}. In both cases a partial wave analysis
was performed, and the decay mode $\rho^{0}\pi^{-}$ was observed.
There is also evidence for $\rhohat(1600)$ in $\eta^{'}\pi$ peaking at 
1.6 GeV \cite{ryabchikov97}.
It has been argued that the $\rho\pi,\;
\eta^{'}\pi$ and $\eta\pi$ couplings of this state qualitatively 
support the
hypothesis that it is a hybrid meson, although other interpretations
cannot be entirely eliminated~\cite{page97exo}.

Recent flux--tube and other model estimates~\cite{bcs} and lattice 
gauge 
theory calculations~\cite{perantonis90} for the lightest
$1^{-+}$ hybrid support a mass substantially higher than 1.4 GeV and 
often 
above 1.6 GeV~\cite{page97exo}. Further, on quite general grounds, it 
can be shown that an $\eta\pi$ decay of $1^{-+}$ hybrids is unlikely
 \cite{page97sel1}. There is 
thus an 
apparent conflict between experimental observation and theoretical 
expectation as far as the 1.4 GeV peak is concerned.

The purpose of the present paper is to propose a resolution of this 
apparent 
conflict. Two possible hypotheses are considered.

\begin{enumerate}

\item The two states are indeed separate resonances and are hybrid 
mesons: the lower one the ground state and the upper one an excited 
state. 
We perform calculations in the flux--tube model of Isgur and Paton
\cite{ip} 
to demonstrate that both on mass and decay grounds, this hypothesis is
implausible.

\item 

We suggest a mechanism whereby an appropriate $\eta\pi$ decay of a
hybrid meson can be generated and argue that there is only one 
$J^{PC}=1^{-+}$ isovector exotic, the lower--mass signal in the E852 
experiment being an artefact of the production dynamics. 
We demonstrate explicitly that is possible
to understand the 1.4 GeV peak observed in $\eta\pi$ as a consequence 
of a 1.6 GeV resonance interfering with a non--resonant Deck--type 
background with an appropriate relative phase.
We do {\it not } propose that there should necessarily be a 
peak at 1.4 GeV; but that if experiment unambiguously confirms
a peak at 1.4 GeV, it can be understood as 
a 1.6 GeV resonance interfering with a non--resonant background. 
 
\end{enumerate}

\section{Hypothesis I: Two hybrid mesons close in mass}

The simplest explanation for the experimental 
report of two peaks at two different masses, is that they are indeed 
separate
Breit--Wigner resonances. 

The most conservative assumption is that these are then both hybrid 
mesons.
Other less likely hypotheses, such as glueball, four--quark and molecular 
interpretations, are discussed in 
ref. \cite{page97exo}. 

In the hybrid scenario, the 1.4 GeV resonance would naturally be assumed to 
be the 
ground state hybrid and the 1.6
GeV resonance an excited hybrid. A numerical calculation in 
the 
flux--tube model indicates
that the orbitally excited D--wave hybrid is the lowest excitation 
above the 
P--wave ground state, with
a mass difference of 400 MeV for light quarks~\cite{bcs}. The same model
predicts $c\bar{c}$ D--wave hybrids to be 270 MeV heavier than the 
ground 
state hybrid~\cite{bcs}, in good agreement with the result of 230 MeV 
found in 
adiabatic--limit lattice--gauge theory simulations~\cite{perantonis90} 
and similar results in NRQCD lattice simulations \cite{collins97}.
Also,
the lattice--gauge calculations find that the next highest levels in the 
$c\bar{c}$ sector are the radially excited P--wave hybrids \cite{collins97},
 which 
are 400 MeV heavier than the ground state~\cite{perantonis90}.
A mass difference of 400 MeV for the light--quark hybrids is clearly 
inconsistent with the
experimental claim of resonances at 1.4 and 1.6 GeV. The absolute mass scale
predicted by theory does not\footnote{Except for a QCD sum rule prediction of $\sim 1.5$ GeV \cite{bal}.} support a ground state hybrid at 1.4 GeV, as
 discussed in ref. \cite{page97exo}. 
Thus there are
two arguments on mass grounds for discarding this hypothesis. 

Further, from the viewpoint of decays, it is qualitatively hard to 
explain 
why the lower--mass $1^{-+}$ hybrids should be seen only in $\eta\pi$. 
This is because relativistic symmetrization selection rules 
suppress 
the $\eta\pi$ decay of 
{\it any} $1^{-+}$ hybrid in QCD in the absense of final state
 interactions~\cite{page97sel1}. 
Within the flux--tube model and constituent--gluon 
models there is a selection rule which suppresses decays of ground 
state 
hybrids to two S--wave mesons~\cite{page95light,pene98}.
This selection rule requires only the standard assumptions of 
non--relativistically moving quarks and
spin 1 pair creation in a connected decay topology~\cite{page97sel2}.
In addition, for $1^{-+}$ hybrids the selection rule is only operative
when the non--relativistic spin of the $Q\bar{Q}$ is 1.
The lowest orbitally excited hybrid in the flux--tube model has $Q\bar{Q}$ 
in spin 1, and 
hence obeys the selection rule. The 
ground and lowest excited hybrids have hence got similar overall decay 
structure.  

\begin{table}[t]
\begin{center}
\caption{Decay widths of a ground state at 1.4 GeV in the flux--tube model. 
The $f_2\pi,\; \rho\omega,\; K^{\ast}K,\; \eta\pi$ and $\eta^{'}\pi$ 
modes are all substantially below 1 MeV.
The conventions and parameters are
those of ref. \protect\cite{page95light}, except for 
the following changes for the $b_1\pi$ and $f_1\pi$ modes. 
Here we use a radial dependence of the hybrid $\sim r$, which 
produces widths $\sim 5\%$ different from ref. \protect\cite{page95light}.
More importantly, we take into account the fact that the $b_1$ and $f_1$
have finite widths, and we assume that they decay predominantly to
$\omega\pi$ and $a_{0}(980)\pi$ respectively.} 
\begin{tabular}{|c|c||r|}
\hline 
Decay Mode & Partial Wave & Width (MeV)  \\
\hline 
$b_1\pi$ & S,D   & 96   \\
$f_1\pi$ & S,D   & 13    \\
$\rho\pi$& P     &  4   \\
$\eta(1295)\pi$ & P & $<1$ \\
\hline 
\end{tabular}
\end{center}
\end{table}

Flux--tube model predictions for the decay of a 
1.4 GeV hybrid are given in Table 1.
We note that the total predicted width of $\sim 110$ MeV is 
much smaller than the observed value.
The calculations show that we expect an appreciably 
larger $\rho\pi$ width than $\eta\pi$ width
for the ground state hybrid.  This is confirmed by QCD sum rule 
calculations \cite{qcdsum}. It then becomes difficult to understand how 
there can be almost no presence of 
$1^{-+}$ wave in the $\rho\pi$ experimental data at 1.5 GeV \cite{bnl97}, 
where there should be significant 
presence due to the $\sim 400$ MeV width of the E852 1.4 GeV state. This 
calls into 
question the interpretation of the 1.4 GeV state as a ground state hybrid.
When final state interactions are taken into account 
(a point on which we elaborate below), we expect a larger $\eta\pi$ width, 
which may invalidate
the preceding arguments. We shall hence proceed with the hypothesis that 
the 1.4 GeV state
is the ground state hybrid and the 1.6 GeV state the orbitally excited 
hybrid.

\begin{table}[t]
\begin{center}
\caption{Decay widths of an orbitally excited hybrid at 1.6 GeV to 
$P+S$--wave states in the flux--tube model in MeV. The conventions and 
parameters are
those of ref. \protect\cite{page95light}. The derivation of the widths
 is discussed in Appendix \protect\ref{appa}. The inverse radius
of the hybrid $\beta_{\rhohat} = 0.27$ GeV is taken to be the same as that
 of the ground state hybrid \protect\cite{page95light}. We also quote an 
error based on taking $\beta_{\rhohat} = 0.23$ GeV. Widths of a 2 GeV
 orbitally excited hybrid is also indicated for comparison.} 
\label{table1}
\begin{tabular}{|c|c||rr|rr|}
\hline 
Decay  & Partial           & \multicolumn{4}{c|}{Excited Hybrid Mass} \\
Mode  & Wave          & \multicolumn{2}{c|}{1.6 GeV} & 
\multicolumn{2}{c|}{2.0 GeV} \\
\hline 
$b_1\pi$ & S   & 118   &   (-22)  &   10    &  (-5)    \\
         & D   &   .1  &          &    .8   &          \\
$f_1\pi$ & S   &  30   &    (-5)  &    4    &  (-1)    \\
         & D   &   .05 &          &    12   &  (-1)    \\
$f_2\pi$ & D   &  .08  &          &    2    &          \\
$a_1\eta$& S   &    -  &          &    11   &  (-3)    \\
         & D   &    -  &          &     2   &          \\
$a_2\eta$& S   &    -  &          &   .07   &          \\
$K_1(1270)K$& S&    -  &          &   105   &  (-27)   \\
         & D   &    -  &          &    .5   &          \\
$K_1(1400)K$&S &    -  &          &   .05   &          \\
         & D   &    -  &          &    .3   &          \\
$K_2^{\ast}(1430)K$& D& -&        &   .03   &          \\
\hline 
\end{tabular}
\end{center}
\end{table}

According to the flux--tube model calculations in Table 2, the 
orbitally excited
hybrid at 1.6 GeV has a somewhat larger total width than the the ground 
state hybrid at 1.4 GeV.  This is a strong 
theoretical statement as generally nodes in orbital wave functions tend to 
suppress
specific partial widths relative to the ground state. Note that P--wave 
modes like $\eta(1295)\pi, \; K^{\ast}K, \; \rho\omega$ should all
be stronger\footnote{In constituent gluon models, the lowest--lying 
excited hybrid is expected to have $Q\bar{Q}$ spin 0, so that
decays to S--wave mesons are not suppressed \cite{pene98,page97sel2}. The 
lowest--lying excited hybrid would then be very wide indeed.} for a 1.6 GeV 
state than a 1.4 GeV state, simply due to phase space. Thus there is a
further problem in understanding why the 1.4 GeV state should have a 
larger experimental width than the 1.6 GeV state.

So on a multiplicity of grounds we are forced to conclude that the
hypothesis that the 1.4 and 1.6 GeV states are both hybrid mesons is 
theoretically untenable.

\section{Hypothesis II: A single hybrid meson at 1.6 GeV}

The current experimental data on the 1.6 GeV state is consistent with mass 
predictions
and decay calculations for a hybrid meson \cite{page97exo,page97had}. This 
then leaves open the
interpretation of the structure at 1.4 GeV. 

There are two basic problems to be solved. Firstly it is necessary to 
find a mechanism which can generate a suitable $\eta\pi$ width for the
hybrid. Then having established that, it is necessary to provide a 
mechanism to
produce a peak in the cross section which is some way below the real
resonance position.
   
We first show that a sizable $\eta\pi$ width for a hybrid 
resonance can be generated by final--state interactions. For this we use 
a doorway calculation, the procedures for which are well established 
\cite{zou94}. We use the simplest approach to provide an upper limit.

The $\eta\pi$ peak in the E852 data spans the $\rho\pi$ and $b_1\pi$ 
thresholds, so
we propose a Deck--type model~\cite{asc} as a source of a non--resonant 
$\eta\pi$ background. 
We then show that, within the K--matrix formalism, interference
between this background and a resonance at 1.6 GeV can
account for the E852 $\eta\pi$ data. The width used for the decay of 
the 1.6 GeV hybrid to $\eta\pi$ is comfortably below the upper limit 
established in the doorway calculation.

\subsection{$\eta\pi$ doorway width of a 1.6 GeV state}

Although the $\eta\pi$ width of a hybrid is suppressed by 
symmetrization selection rules \cite{page97sel1} which
operate on the quark level and have been estimated in QCD sum rules 
to be tiny ($\sim$ 0.3 MeV) \cite{qcdsum},
long distance contributions to this width are possible. 
We shall show that these can be very much larger than the widths 
obtained without final state interactions. 

\begin{figure}
\begin{center}
\caption{\plabel{fig1} Decay of $\rhohat$ to $\eta\pi$ via final state 
interactions.}
\leavevmode
\hbox{\epsfxsize=5 in}
\epsfbox{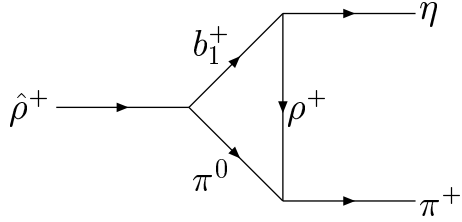}
\end{center}
\end{figure}

The procedure we adopt is that of a doorway calculation with on--shell
mesons \cite{zou94}, which provides an upper limit. An essential 
ingredient is the
presence of an allowed dominant decay which can couple strongly to the 
channel of interest. In the flux--tube model $b_1\pi$ is such
a dominant decay, and it is strongly coupled to $\eta\pi$ by $\rho$ 
exchange (see Figure \ref{fig1}).
So we consider the process $\rhohat^+ \rightarrow b_1\pi \rightarrow 
\eta\pi^+$. 

For on--shell states the Lorentz invariant amplitudes can be 
parameterized as

\beqn {\cal M}_{\rhohat\rightarrow b_1^+\pi^0} = \epsilon_{\mu}^
{\rhohat}
\epsilon_{\nu}^{b_1\ast}(\hs g^{\mu\nu} +\hd p^{\mu}_{b_1} p^{\nu}_
{\rhohat})
\eeqn

\beqn {\cal M}_{b_1\rightarrow \rho^+\eta} = \epsilon_{\mu}^{b_1}
\epsilon_{\nu}^{\rho\ast}(\fs g^{\mu\nu} +\fd p^{\mu}_{\rho} p^{\nu}_
{b_1})
\eeqn

\beqn \plabel{rhopipi} {\cal M}_{\rho\rightarrow \pi^+\pi^0}  = \ggg 
\epsilon_{\mu}^{\rho}(p^{\mu}_{\pi^+}-p^{\mu}_{\pi^0})\eeqn

where $p_X^{\mu}$ and $\epsilon^X_{\mu}$ refer to the momentum and 
polarization 4--vectors of X respectively, $g^{\mu\nu}$ is the flat 
space metric tensor, and $\hs,\; \hd,\; \fs,\; \fd $ and $\ggg$ are decay 
constants to be determined. These are discussed in Appendix B.

The doorway amplitude for the process $\rhohat \rightarrow b_1^+\pi^0
\rightarrow \eta\pi^+$ is then

\beqna \lefteqn{{\cal M}_{doorway} = \frac{i}{(2\pi)^4}\int 
d^4 p_{b_1}\;
\epsilon_{\mu}^{\rhohat}(\hs g^{\mu\nu} +\hd p^{\mu}_{b_1} p^{\nu}_
{\rhohat})
\frac{g_{\nu\sigma}}{p_{b_1}^2-m_{b_1}^2+i\epsilon} \nonumber } \\ & &
\times \; (\fs g^{\sigma\lambda} +\fd p^{\sigma}_{\rho} p^{\lambda}_
{b_1})
\frac{g_{\lambda\kappa}}{p_{\rho}^2-m_{\rho}^2+i\epsilon} 
\ggg (p^{\kappa}_{\pi^+}-p^{\kappa}_{\pi^0})
\frac{1}{p_{\pi^0}^2-m_{\pi^0}^2+i\epsilon} \eeqna

where $\epsilon$ is a small real number. Here we have contracted the 
Lorentz 
indices on the internal vector particles in the usual way \cite{zou94}, 
effectively working in the ``Feynman gauge''. Integration is performed
over 
the loop momentum. We evaluate the doorway amplitude in the rest 
frame of
$\rhohat$ using conservation of momentum at the vertices and the Cutkovsky 
rules to obtain \cite{zou94}

\beqn {\cal M}_{doorway} = \frac{i \ggg} {32\pi (E_{b_1}+E_{\pi^0})q}
\;\{ (\cc+\dd z+\ee z^2) \ln|\frac{1+z}{1-z}|-2(\dd+z\ee) \}
\eeqn

if the component of the polarization of $\rhohat$ is in the direction 
of the 
outgoing particles, i.e. $\eta$ or $\pi^+$. For other polarizations, 
${\cal M}_{doorway} = 0$. 

Here

\beqn \cc=-\hs q \;\{\fs + \fd (E_{b_1}(E_{\pi^0}+E_{\pi^+})+p^2)\} \eeqn

\beqna \lefteqn{\dd=p\;\{ -\hs \fs + \hd \fs m_{\rhohat} (E_{\pi^0}+
E_{\pi^+}) + 
 \nonumber } \\ & & (\hs \fd + \hd \fd m_{\rhohat} (E_{b_1}-E_{\eta}))
(E_{b_1}(E_{\pi^0}+E_{\pi^+})+p^2) - \hs \fd q^2\} \eeqna

\beqn \ee = \fd p^2 q \;\{ \hs + \hd m_{\rhohat}(E_{b_1}-E_{\eta}) \} \eeqn

\beqn z = \frac{p^2+q^2-(E_{b_1}-E_{\eta})^2+m_{\rho}^2}{2 p q}\eeqn

and $p$ is the magnitude of the $b_1$ or $\pi^0$ momentum, $q$ the 
magnitude of the
$\eta$ or $\pi^+$ momentum, and $E_X$ the energy of X, all in the rest
frame of $\rhohat$. 

The doorway decay width for $\rhohat^+\rightarrow b_1\pi\rightarrow 
\eta\pi^+$ is then

\beqn \plabel{doorwid} \Gamma_{\rhohat\rightarrow b_1\pi\rightarrow 
\eta\pi^+} = \frac{q}{8\pi m_{\rhohat}^2}\;\frac{1}{3}\; |2 \; 
{\cal M}_{doorway}|^2 
\eeqn

where we have taken into account that there are two possible 
intermediate 
processes contributing to the total amplitude, i.e. $\rhohat^+ 
\rightarrow 
b_1^+\pi^0 \rightarrow \eta\pi^+$ and $\rhohat^+ \rightarrow b_1^0\pi^+ 
\rightarrow \eta\pi^+$.

We calculate that the doorway width is
$57\pm 14$  MeV for all the particles on--shell, taking into account 
uncertainties in the couplings 
$b_1^+\rightarrow\rho^+\eta$ and $\rho^+\rightarrow\pi^+\pi^0$ 
(see Appendix \ref{appb}).
As remarked in Appendix \ref{appb}, there are uncertainties in the coupling
$\rhohat^+\rightarrow b_1^+\pi^0$ which can make this doorway width up to 
$\sim 40\%$ smaller. 
Thus we conclude that the doorway width is less than $57\pm 14$ MeV.
It should also be remembered that the doorway calculation as
it stands provides an upper limit, since we would get a smaller answer 
if we were
to take one of the internal legs off--shell and introduce form factors 
\cite{gort}. However these are unknown, and as $57\pm 14$ MeV is well above
the $\eta\pi$ width required this does not create a problem.

\subsection{Non--resonant $\eta\pi$ Deck background} 

The 1.4 GeV peak in the $\eta\pi$ channel occurs in the vicinity of
the $\rho\pi$ and $b_1\pi$ thresholds, and it is therefore natural
to consider these as being responsible in some way for the $\eta\pi$
peak. The Deck mechanism~\cite{asc} is known to produce broad low--mass
enhancements for
a particle pair in three--particle final states, for example in $\pi p 
\rightarrow (\rho\pi) p$. In this latter case, the incident pion 
dissociates into $\rho\pi$, either of which can then scatter off the 
proton~\cite{stod}. At sufficiently high energy and presumed dominance 
of the exchange of vacuum
quantum numbers (pomeron exchange) for this scattering one obtains the
``natural parity change'' sequence $\pi \rightarrow 0^-, 1^+, 2^-....$
(the Gribov--Morrison rule~\cite{gm}). However if the scattering 
involves the exchange of
other quantum numbers then additional spin-parity combinations can be
obtained, including $J^P = 1^-$. This can be seen explicitly in 
ref. \cite{asc} for the reaction $\pi p \rightarrow (\rho\pi) p$
in which the full $\pi p$ scattering amplitude was used, so that
the effect of exchanges other than the pomeron are automatically
included. The $J^P$ sequence
from the ``natural parity change'' dominates due to the dominant 
contribution 
from pomeron exchange, but other spin-parity states are present at a 
non--negligible level. The Reggeised Deck effect can simulate resonances, 
both in terms of the mass distribution and the phase \cite{asc,bow}. 
It can produce 
circles in the Argand plot, the origin of which is the Regge phase factor
exp$[-i{{1}\over{2}}\pi\alpha(t_R)]$.  

\begin{figure}[t]
\begin{center}
\caption{\plabel{fig2} Deck background production in $\eta\pi$.}
\leavevmode
\hbox{\epsfxsize=5 in}
\epsfbox{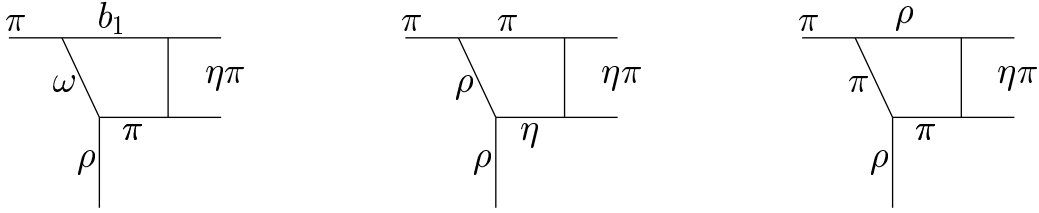}
\end{center}
\end{figure}

It is also important to note that rescattering of the lighter particle 
from the dissociation of the incident beam particle is not a 
prerequisite, and indeed
both can contribute~\cite{stod}. We suggest that in our particular 
case the
relevant processes are (from left to right in Figure \ref{fig2})

\begin{enumerate}

\item $\pi \rightarrow b_1\omega$, $\omega p \rightarrow \pi p$ giving a
$b_1\pi$ final state.
\item $\pi \rightarrow \pi\rho$, $\rho p \rightarrow \eta p$ giving a
$\eta\pi$ final state.
\item $\pi \rightarrow \rho\pi$, $\pi p \rightarrow \pi p$ and $\rho p
\rightarrow \rho p$ giving a $\pi\rho$ final state. 

\end{enumerate}

For each of these processes the rescattering will be predominantly via
$\rho$ (natural parity) 
exchange to give the required parity in the final state.
Obviously process (ii) produces a final $\eta\pi$ state directly, but 
for (i) and (iii) the $b_1\pi$ and $\rho\pi$ final states are required
to rescatter into $\eta\pi$ (for which the doorway calculation provides
an explicit mechanism).

Unfortunately only the $\pi p$ cross section can be obtained with any
reliablity. The others can be estimated with varying degrees of 
uncertainty from:

\begin{enumerate}
\item $\pi\rho\rightarrow b_1\pi$: data on $\pi^0 p\rightarrow\omega 
n$, which can be inverted to give $\omega p\rightarrow \pi^0 p$.
\item $\pi\rho\rightarrow\eta\pi$: data on $\gamma p\rightarrow\eta 
p$, which by assuming vector meson dominance can give 
$\rho p\rightarrow\eta p$.
\item $\pi\rho\rightarrow\rho\pi$: data on $\pi^- p\rightarrow\pi^0 
n$ and on $\pi^{\pm} p \rightarrow \pi^{\pm} p$; data on $\gamma p
\rightarrow \rho p$ and vector meson dominance. 
\end{enumerate}

In view of the uncertainties in the underlying
reactions, the lack of an explicit value for the $J^{PC} = 1^{-+}$
$\eta\pi$ cross section in the E852 experiment, and the impossibility
of a precise evaluation of the rescattering into the $\eta\pi$ channel
from $\rho\pi$ and $b_1\pi$, we have not attempted a complete Deck--type
calculation. We concentrate rather on the mass--dependence which it
generates. The characteristic mass--dependence is a peak just above
the threshold. Thus there are three peaks from our proposed mechanism:
a sharp peak just above the $\eta\pi$ threshold; a broader one at about
1.2 GeV from the $\rho\pi$ channel; and a very broad one at about 1.4
GeV from the $b_1\pi$ channel. The first of these is effectively
removed by experimental cuts, but the net effect of the two latter is 
to produce a broad peak in the $\eta\pi$ channel. Thus invoking this 
mechanism does provide an explanation of the larger width of the $\eta\pi$ 
peak at 1.4 GeV in the E852 data compared to that of the $\rho\pi$ peak at 
1.6 GeV. Because of the resonance--like nature of Deck amplitudes it is also
possible in principle to simulate the phase variation observed. However 
as there are Deck amplitudes and the 1.6 GeV resonance, presumably 
produced directly, it is necessary to allow for interference between them. 
We use the K--matrix formalism to calculate this, and also to
demonstrate that the Deck mechanism is essential to produce the 1.4 GeV
peak.

\subsection{K--matrix with P--vector formalism}

It is straightforward to demonstrate that within the K--matrix 
formalism it is impossible 
to understand the $\eta\pi$ peak at 1.4 GeV as due to a 1.6 GeV state
if only resonant decays to $\eta\pi$, $\rho\pi$ and $b_1\pi$ are 
allowed despite the strong threshold effects in the two latter channels
\footnote{The use of $b_1\pi$ is not critical
here: any channel with a threshold near 1.4 GeV will suffice.}.
We find that for a $b_1\pi$ width of $\approx 200$ MeV and $\eta\pi$ and
$\rho\pi$ widths in the region $1-200$ MeV there is no shift of the peak.
However, when a non--resonant $\eta\pi$ P--wave is introduced, the 
interference between this and the 1.6
GeV state can appear as a 1.4 GeV peak in $\eta\pi$.
 
We have seen that the non--resonant $\eta\pi$
wave can have significant presence at the $b_1\pi$ or $f_1\pi$ threshold 
(called the
``P+S'' threshold), e.g. $1.368$ GeV for $b_1\pi$, because
of the substantial ``width'' generated by the Deck mechanism. Since the 
hybrid is believed to couple 
strongly to ``P+S'' states due to selection rules \cite{page95light,
page97sel2}, 
the interference effectively shifts the peak in 
$\eta\pi$ down from 1.6 GeV to 1.4 GeV. It is not necessary for the 1.6
GeV resonance to have a strong $\eta\pi$ decay. It is significant that the
E852 experiment
finds $\rhohat$ at $1370\pm 16^{+50}_{-30}$ MeV, near the $b_1\pi$ 
threshold, but not at 1.6 GeV.  
It is possible for a state to peak near the 
threshold of the channel to which it has a strong coupling, assuming
that the (weak) channel in which it is observed has a significant 
non--resonant origin. 

We follow the K--matrix formalism in the P--vector approach as 
outlined in \cite{ait,suhurk}. We assume there to be a $\rhohat$ with
$m_{\rhohat} = 1.6$ GeV  as motivated by the structure 
observed in $\rho\pi$ \cite{bnl97}. The problem is simplified to the 
case where there is decay to two observed channels i.e $\eta\pi$ and
$\rho\pi$, and one unobserved $P + S$ channel. These channels are
denoted 1, 2 and 3 respectively. The production amplitudes and the
amplitude after final--state interactions are grouped together in 
the 3-dimensional P-- and F--vectors respectively. In order to
preserve unitarity \cite{ait} we assume a real and symmetric 
$3 \times 3$ K--matrix. The amplitudes after final--state interactions
and production are related by \cite{ait}  

\beqn  F = (I-iK)^{-1} P \eeqn

We define the widths as

\beqn\plabel{gam} \Gamma_i = \gamma_i^2 \;\Gamma_{\rhohat} \frac{B^2(q_i)}
{B^2(q_i^{\rhohat})}\rho(q_i)\hspace{1cm} i=1,2 \eeqn

\beqn \Gamma_3 = \gamma_3^2 \;\Gamma_{\rhohat}\; \rho(q_3) \eeqn

where $q_i$ is the breakup momentum in channel $i$ from a state of 
effective mass $w$, and $q_i^{\rhohat}$ is the breakup momentum in channel 
$i$ from a state of effective mass $m_{\rhohat}$. The kinematics is 
taken care of by use of the phase space factor

\beqn \rho(q)=\frac{2 q}{w}\eeqn

and the P--wave angular momentum barrier factor

\beqn  B^2(q) = \frac{(q/q_R)^2}{1+(q/q_R)^2}\eeqn

where the range of the interaction is $q_R = 1$ fm $= 0.1973$ GeV.

We assume the experimental width in $\rho\pi$ of $\Gamma_{\rhohat}=168$
 MeV \cite{bnl97} to be the total width of the state\footnote{It is 
found that our results in Fig. 3 are very similar even for a 
width of 250 MeV.}.
We adopt the flux--tube model of 
Isgur and Paton \cite{ip} and
use the $\rho\pi$ and $b_1\pi$ widths which it predicts for a hybrid
of mass 1.6 GeV. Since the model predicts that the branching ratio
of a hybrid to $b_1\pi$ is $59-74$ \% and to $f_1\pi$ is $12-16$ \%
\cite{page97had}, we obtain the $P+S$--wave width to be $120-150$ MeV.  
Analysis of the data
shows that the $\rho\pi$ branching ratio of $\rhohat(1600)$ is $20\pm 2$ \%
\cite{page97exo}, corresponding to a $\rho\pi$ width of 
$30-37$ MeV.  This is consistent with flux--tube model predictions
of $9-22$ \% \cite{page97had}. For the simulation we use a $b_1\pi$ width
of 120 MeV, a $\rho\pi$ width of 34 MeV, and an $\eta\pi$ width of 14 MeV,
well within the limits set by the doorway calculation. We neglect other 
predicted modes of
decay since we restrict our analysis to three channels.

The K--matrix elements are 

\beqn \plabel{kmat} K_{ij} = \frac{m_{\rhohat}\sqrt{\Gamma_i \Gamma_j}}
{m_{\rhohat}^2-w^2} + c_{ij}\eeqn

where $c_{ij}$ includes the possibility of an unknown background.

In the simulation we assume that the Deck terms can be treated as 
conventional resonances.
This is not necessary, but is done to reduce the number of free parameters. 
We assume that the $\eta\pi$ Deck amplitude is produced predominantly via 
the $b_1\pi$ and $\rho\pi$ channels, and so is modelled as a resonance
 at a mass $m_{b1} =
1.32$ GeV and a width $\Gamma_{b1} = 300$ MeV. This width fits
the E852 data at low $\eta\pi$ invariant masses (see Figure \ref{3ch}a).
The $\rho\pi$ background is assumed
to peak at a mass $m_{b2} = 1.23$ GeV with a width $\Gamma_{b2} = 400$ MeV, 
which when plotted as an invariant mass distribution effectively peaks
at $\sim 1.15$ GeV, in agreement with detailed Deck calculations 
in the $1^{++}$ wave \cite{asc}.

We incorporate the $\eta\pi$ and $\rho\pi$ Deck background by putting 
$c_{ij}=0$ except for

\beqn c_{11}=\frac{m_{b1}\Gamma_{b1}}{m_{b1}^2-w^2} \hspace{2cm}
c_{22}=\frac{m_{b2}\Gamma_{b2}}{m_{b2}^2-w^2}
\eeqn

The widths 
are defined analogously to Eq. \ref{gam} as

\beqn \Gamma_{bi} = \gamma_{bi}^2 \;\Gamma_{\rhohat}\; \frac{B^2(q_i)}
{B^2(q_i^{b})}\rho(q_i)\hspace{1cm} i=1,2  \eeqn

where $q_i^b$ is the breakup momentum from a state of effective mass 
$m_{bi}$ (for $i=1,2$).

The production amplitudes are given by

\beqn P_i = \frac{m_{\rhohat} V_{\rhohat} \sqrt{\Gamma_i 
\Gamma_{\rhohat}}}{m_{\rhohat}^2-w^2} + c_i \eeqn

where the (dimensionless) complex number $V_\rhohat$ measures the 
strength of the production of $\rhohat$. We take $c_3=0$ and 

\beqn
c_1 = \frac{m_{b1} V_{b1} \sqrt{\Gamma_{b1} 
\Gamma_{\rhohat}}}{m_{b1}^2-w^2} \hspace{2cm}
c_2 = \frac{m_{b2} V_{b2} \sqrt{\Gamma_{b2} 
\Gamma_{\rhohat}}}{m_{b2}^2-w^2} 
\eeqn
where the complex numbers $V_{bi}$ gives the production strengths of the 
Deck background in channel $i$.  

The results of this fit are shown in Fig. \ref{3ch} and clearly provide a 
good
description of the $\eta\pi$  data \cite{bnletapi,suhurk}. 

We briefly discuss the results.
Fig. \ref{3ch}a indicates a steep rise for low invariant $\eta\pi$ masses, 
and a slow fall for 
large $\eta\pi$ masses. This naturally occurs because of the presence of 
the resonance at 1.6 GeV in the high mass region, which shows as a 
shoulder in our fit. 
Figure \ref{3ch}b reproduces the experimental slope and phase change
in $\eta\pi$ \cite{suhurk}. One might find this unsurprising, since the
background changes phase like a resonance. However, we have confirmed,
by assuming a background that has constant phase as a function of
$\eta\pi$ invariant mass, that the experimental phase shift is still
reproduced. The experimental phase shift is hence induced by the 
resonance at 1.6 GeV.


Our fit to E852 $\eta\pi$ and $\rho\pi$ data (with a prediction for the
$b_1\pi$ data) requires 12 independent parameters (see the caption of 
Figure 3).

Without the inclusion of a dominant $P+S$--wave channel
\footnote{The $b_1\pi$ coupling of the resonance is set to zero, with
the $\eta\pi$ and $\rho\pi$ couplings the same as before.} the $\eta\pi$ 
event shape
clearly shows two peaks, one at 1.3 GeV and one at 1.6 GeV, which is not 
consistent
with the data \cite{bnletapi}. The phase motion is also more pronounced in 
the 
region between the two peaks than that suggested by the data \cite{suhurk}.
The r\^{o}le of the dominant $P+S$--channel is thus that at invariant 
masses between
the two peaks, the formalism allows coupling of the strong $P+S$ channel 
to $\eta\pi$, so that the
$\eta\pi$ appears stronger than it would otherwise, interpolating between 
the peaks at
1.3 and 1.6 GeV, consistent with the data \cite{bnletapi}. A dominant $P+S$
 decay of the $\rhohat$
is hence suggested by the data. 

\section{Discussion}

We have argued that on the basis of our current understanding of meson
masses it is implausible to interpret the 1.4 GeV peak seen in the
$J^{PC} = 1^{-+}$ $\eta\pi$ channel by the BNL E852 experiment as 
evidence for an exotic resonance at that mass. We acknowledge that
this is not a proof of non--existence and note the Crystal Barrel
claim for the presence of a similar state at $1400 \pm 20\pm 20$ MeV in
the reaction $p\bar p \rightarrow \eta\pi^+\pi^-$. However this is
not seen as a peak and is inferred from the interference pattern
on the Dalitz plot. It has not been observed in other channels in
$p\bar p$ annihilation at this mass, which is required for
confirmation. So at present we believe that the balance of 
probability is that the structure does not reflect a real resonance.

Given this view, it is then necessary to explain the data and in
particular the clear peak and phase variation seen by the E852 
experiment. Additionally the observation of the peak only in the
$\eta\pi$ channel, which is severely suppressed 
by symmetrization selection rules, requires justification. We have dealt with
these two questions in reverse order. We first demonstrate that
final--state interactions can generate a sizable $\eta\pi$ decay. 
We believe that this result by itself is
of considerable significance and is of wider relevance.
We then suggest that the E852 $\eta\pi$ peak is due
to the interference of a Deck--type background with a hybrid
resonance of higher mass, for which the $\rhohat$ at 1.6 GeV
is an obvious candidate. This mechanism also provides the
natural parity exchange for the former which is observed experimentally.
The parametrization of the Deck background is found not to be critical.


A key feature in our scenario is the presence of the large ``$P+S$''
amplitude which drives the mechanism. This should be observable both
as a decay of the 1.6 GeV state and as a lower--mass enhancement due to
the Deck mechanism. Depending on the relative strength of these two 
terms the resulting mass distribution could be considerably distorted
from a conventional Breit--Wigner shape as the Deck peak is broad and
the interference could be appreciably greater than in the $\rho\pi$ channel.

\vspace{.9cm}                                            

\noindent {\bf Acknowledgements}

\vspace{.2cm}

Helpful discussions with S.-U. Chung and D.P. Weygand are acknowledged.
The work of A.D. was supported by the University of Manchester and an 
invitation from TJNAF. 
P.R.P. acknowledges a Lindemann Fellowship from the English Speaking 
Union.

\appendix

\section{Appendix A: Decay of orbitally excited hybrid \plabel{appa}}

We detail here the flux--tube model calculation for the decay of
an orbitally excited hybrid to $P+S$--wave mesons. The normalized wave
 functions
of the P--wave and S--wave mesons are just S.H.O. wave functions with 
the same inverse radius $\bt$ \cite{page95light}, and are respectively

\beqn\plabel{mewav}
2 \sqrt{\frac{2}{3}}\; \frac{\bt^{5/2}}{\pi^{1/4}}\; r Y_{1 M_{L}}
 (\hat{\br})\exp - \frac{1}{2}{\bt^{2}r^{2}}
\hspace{3em}   \frac{\bt^{3/2}}{\pi^{3/4}}\; \exp - \frac{1}{2}{\bt^{2}r^{2}}
\eeqn

The normalized
wave function of the orbitally excited hybrid is taken to be

\beqn \plabel{hywav} {\cal N}_{\rhohat} r^\delta {\cal D}_{M^{\rhohat}_L
\Lambda}^L(\phi,\theta,-\phi)\exp -\frac{1}{2}\beta_{\rhohat}^2r^2
\hspace{1cm}{\cal N}_{\rhohat} =  \sqrt{\frac{(2L+1)\beta_{\rhohat}^
{3+2\delta}}{2\pi\Gamma(\frac{3}{2}+\delta)}}\eeqn

where the Wigner D--function ${\cal D}_{M^{\rhohat}_L\Lambda}^L$ guarantees 
that the state has
total orbital angular momentum $L=2$, the first orbital excitation above 
the ground state with total orbital angular momentum 1,
$M^{\rhohat}_L$ is the total orbital angular momentum projection, and 
$\Lambda=\pm 1$ is the angular momentum of the flux--tube
around the $Q\bar{Q}$--axis \cite{ip}. The inverse radius 
$\beta_{\rhohat}$ characterizes the size of the wave function and $\Gamma$ 
is the Gamma--function. The radial dependence is chosen to be proportional 
to $r^{\delta}$, where $\delta$ is chosen such 
radial Schr\"{o}dinger equation \cite{ip} in the limit $r\rightarrow 0$, 
which leads to the condition $\delta(\delta+1)=L(L+1)-\Lambda^2$
\cite{page95light}, implying that $\delta=1.79$ for $L=2$. The lowest 
orbitally excited hybrid has the $Q\bar{Q}$ in spin 1, just like the ground 
state hybrid.

The relevant overlap can be obtained by inserting the spacial wave
 functions into the decay matrix element and
performing the integration over the quark--antiquark pair creation
 position \cite{page95light,ip}

\beqna \plabel{interg}
\lefteqn{\hspace{-1.8cm}{\cm}_{M^{\rhohat}_LM_L} =\mbox{flavour}\;
 \frac{0.62\; \gamma_0}{(1+\frac{0.1}{\bt^2})^2}\;  \frac{\sqrt{2}}{\bt}
  {\cal N}_{\rhohat}
\int_0^{2\pi} d\phi\; \int_0^{\pi}\sin\theta\; d\theta\; 
\int_0^{\infty}r^2\; dr\; r^\delta  \;(-i\bt^2\br+\bp)\cdot 
{\bf e}^{\ast}_{M^{\rhohat}_L-M_L}  \nonumber } \\ & & \times\;
{\cal D}_{M^{\rhohat}_L 1}^2(\phi,\theta,-\phi)\;
{\cal D}_{M_L 1}^{1\;\ast}(\phi,\theta,-\phi)\;
\exp \frac{i}{2}\bp\cdot\br\;\exp-\frac{1}{4}r^2(2\beta_{\rhohat}^2+\bt^2)
\eeqna
where $\bp$ is the momentum of the outgoing mesons in the rest frame
 of the hybrid and ${\bf e}$
a spherical basis vector. 
Notice that the pair creation constant $\gamma_0$ of the $^3P_0$ model
enters explicitly in Eq. \ref{interg}. This is because the flux--tube
model, within the assumptions made for the wave functions, gives a
prediction for the couplings of a hybrid in terms of couplings for
mesons in the $^3P_0$ model \cite{page95light} (the constants
0.62 and 0.1 in Eq. 6 are derived from flux--tube dynamics). We take 
$\gamma_0= 0.39$ \cite{page95light,geiger94,ip}. 
The integral in Eq. \ref{interg} is performed numerically.

We can write the decay amplitudes in terms of the amplitudes in Eq.
 \ref{interg} as follows, following ref. \cite{page95light}. 
For $b_1\pi$ (``flavour'' = 2):

\beqna  \plabel{df1}
\lefteqn{\mbox{S--wave amplitude} = - \frac{1}{\sqrt{15}} \Im\;
  (\sqrt{6}\cm_{21} + \sqrt{3} \cm_{11} + \cm_{01} + \sqrt{3} \cm_{10}
 + \cm_{00})\nonumber } \\ & & 
\hspace{-.75cm}\mbox{D--wave amplitude} = - \frac{1}{\sqrt{30}} \Im\;
 (\sqrt{6} \cm_{21} + \sqrt{3} \cm_{11} + \cm_{01} -2 \sqrt{3}
 \cm_{10} -2 \cm_{00})
\eeqna
where $\Im $ selects the imaginary part of the amplitude.
For $f_1\pi$ and $a_1\eta$ (``flavour'' = $\sqrt{2}$ and $1$ respectively):

\beqna  \plabel{df2}
\lefteqn{\mbox{S--wave amplitude} =   \frac{1}{\sqrt{30}} \Im\;
  (\sqrt{6}\cm_{21} + \sqrt{3} \cm_{11} + \cm_{01} + \sqrt{3}
 \cm_{10} + \cm_{00})\nonumber } \\ & & 
\hspace{-.75cm}\mbox{D--wave amplitude} =   \frac{1}{2\sqrt{15}}
 \Im\; (\sqrt{6} \cm_{21} -2\sqrt{3} \cm_{11} -5 \cm_{01} + \sqrt{3}
 \cm_{10}  +\cm_{00})
\eeqna

The $K_1(1270)$ is regarded as $\cos\tilde{\theta}\; |^1P_1\rangle +
 \sin\tilde{\theta}\; |^3P_1\rangle$ and $K_1(1400)$ the orthogonal
 partner, with $\tilde{\theta}=-34^o$ \cite{page95light,ip}. $^1P_1$
 and $^3P_1$ are the P--wave mesons with $Q\bar{Q}$ 
combinations of the decay amplitudes to the $^1P_1$ meson (Eq. \ref{df1})
 and $^3P_1$ meson (Eq. \ref{df2}). For $K_1(1270)K$ and $K_1(1400)K$
 ``flavour'' = $\sqrt{2}$.

For $f_2\pi$, $a_2\eta$ and $K_2^{\ast}(1430)K$ (``flavour'' = $\sqrt{2}$,
 $1$ and $\sqrt{2}$ respectively):

\beqn \plabel{df3}\mbox{D--wave amplitude} = \frac{1}{2\sqrt{5}} \Im\;
 (\sqrt{6} \cm_{21} - \cm_{01} - \sqrt{3} \cm_{10} - \cm_{00}) \eeqn

The decay amplitudes in Eqs. \ref{df1} - \ref{df3} are then used to 
calculate widths according to the phase space conventions of Eq. 6 of
 ref. \cite{page95light}.

\section{Appendix B: Doorway calculation constants \plabel{appb}}

\subsection{$\hs$ and $\hd$}

These couplings cannot be obtained from experiment, as there is 
currently no published data on the $\rhohat^+$ coupling to 
$b_1^+\pi^0$. We have calculated $\hs$ and $\hd$ in the 
non--relativistic flux--tube model of Isgur and Paton, following 
the conventions and methods of ref. \cite{page95light} (except that 
we assume the relativistic phase space convention \cite{geiger94}). 
The wave functions of the mesons and the hybrid are given  respectively
 by Eqns. \ref{mewav} and
\ref{hywav} (with $L=1$).
We find that

\beqn \plabel{flux}\left( \barr{c} \hs \\ \hd \earr \right) = - 
\sqrt{4 m_{\rhohat} E_{b_1} E_{\pi^0}}
\left( \barr{c} -2\qa+3\qd+\qe   \\ \frac{m_{b_1}}{m_{\rhohat}p^2} (3(
\qd+\qe)+ (\frac{E_{b_1}}{m_{b_1}}-1)(-2\qa+3\qd+\qe)) \earr \right)
\eeqn

where

\beqn \left( \barr{c} \qa \\ \qd \\ \qe  \earr \right) = \frac{0.62\;
\gamma_0}{(1+\frac{0.1}{\beta^2})^2}\frac{2}{\beta}\sqrt{\frac{\pi
\beta_{\rhohat}^{3+2\delta}}{3\; \Gamma(\frac{3}{2}+\delta)}} \int_0^{
\infty} dr\; r^{2+\delta} \exp(-\frac{r^2}{4}(2\beta_{\rhohat}^2+
\beta^2)) \left( \barr{c} \beta^2 r j_0(\frac{pr}{2}) \\
 p j_1(\frac{pr}{2}) \\ \beta^2 j_2(\frac{pr}{2}) \earr \right) \eeqn

where $j_i$ and $\Gamma$ refers to the spherical Bessel and Gamma 
functions respectively.
We use 
$\gamma_0=0.53$ which reproduces
conventional meson decay phenomenology for relativistic phase space
 \cite{geiger94}.
In Eq. \ref{flux}, $\beta$ refers to the inverse radius of the $b_1^+$ 
or $\pi^0$,
the parameter that enters in the S.H.O. wave function. Similarly, 
$\beta_{\rhohat}$
is the inverse radius of $\rhohat^+$.

Setting $\delta=1$, $\beta_{\rhohat}=0.27$ GeV and $\beta=0.36$ GeV, 
yields

\beqn \hs = 3.0\; \mbox{GeV}\hspace{1cm}  \hd = -8.2\; \mbox{GeV}^{-1}
\eeqn

We chose a value of $\gamma_0$ towards the upper end of the range in the 
literature.
In calculations of excited mesons, values of $\gamma_0$ as low as .4 have
 been used \cite{biceps}.
The values for $\hs$ and $\hd$ can hence be only $.4/.53$ of the values
 quoted.

\subsection{$\fs$ and $\fs$}

Although the $b_1^+$ coupling to $\rho^+\eta$ is not known 
experimentally \cite{pdg96}, its coupling to $\omega\pi^+$ is well 
known \cite{pdg96}, and can be used to obtain the $\rho^+\eta$ 
coupling. We first derive the $\omega\pi$ coupling by assuming that 
100\% of the decays of $b_1^+$ is to $\omega\pi^+$ and using the 
experimentally measured D--wave to S--wave amplitude ratio.

The amplitude for $b_1^+\rightarrow \omega\pi^+$ can be written as

\beqn {\cal M}_{b_1\rightarrow \omega\pi} = \epsilon_{\mu}^{b_1}
\epsilon_{\nu}^{\omega\ast}(\fsw g^{\mu\nu} +\fdw p^{\mu}_{\omega} 
p^{\nu}_{b_1})\eeqn

Using the Jacob--Wick formulae we write the S--wave and D--wave 
decay amplitudes  as

\beqn  {\cal M}_{b_1\rightarrow \omega\pi}^S = \sqrt\frac{4}{3}
{\cal M}_{b_1\rightarrow \omega\pi}^{\epsilon=\uparrow} + 
\sqrt\frac{1}{3}{\cal M}_{b_1\rightarrow \omega\pi}^{\epsilon=
\rightarrow} = \frac{1}{\sqrt{3}}(-2\fsw-\frac{E_\omega}{m_\omega}
\fsw+p_\omega^2 \frac{m_{b_1}}{m_\omega}\fdw) \eeqn

\beqn  {\cal M}_{b_1\rightarrow \omega\pi}^D = \sqrt\frac{2}{3}
{\cal M}_{b_1\rightarrow \omega\pi}^{\epsilon=\uparrow} - 
\sqrt\frac{2}{3}{\cal M}_{b_1\rightarrow \omega\pi}^{\epsilon=
\rightarrow} = \sqrt\frac{2}{3}((\frac{E_\omega}{m_\omega}-1)\fsw-
p_\omega^2 \frac{m_{b_1}}{m_\omega}\fdw) \eeqn

where all energies and momenta refer to the $b_1^+$ rest frame. 
Relating the S--wave and D--wave amplitudes to the corresponding 
widths in the usual way (analogous to Eq. \ref{doorwid}), we finally 
obtain

\beqn \left( \barr{c} \fsw \\ \fdw \earr \right)  = 2 \sqrt{\frac{\pi}
{p_\omega}}\sqrt{\frac{\Gamma_{total}}{1+(\sqrt\frac{\Gamma^D}
{\Gamma^S})^2}} \left( \barr{c} m_{b_1} (\sqrt{2} + \sqrt\frac
{\Gamma^D}{\Gamma^S} )  \\ \frac{m_{\omega}}{p_\omega^2} (\sqrt{2} 
(\frac{E_\omega}{m_\omega}-1) + (\frac{E_\omega}{m_\omega}+2)\sqrt
\frac{\Gamma^D}{\Gamma^S} ) \earr \right) \eeqn

Using the experimental data $\Gamma_{total} = 142\pm 8$ GeV and 
$\sqrt\frac{\Gamma^D}{\Gamma^S} = + 0.29\pm 0.04$ (where the sign is 
taken to mean that the D--wave and S--wave amplitudes have the same 
sign), we obtain can then deduce the coupling constants

\beqn \fsw = 4.6\pm 0.2\; \mbox{GeV}  \hspace{1cm} \fdw = 14.4 \pm 2.2
\; \mbox{GeV}^{-1}  \eeqn

To obtain the $b_1$ coupling to $\rho^+\eta$, we note that 
(neglecting effects due to phase space), it should be related to the 
$\omega\pi^+$ coupling by a simple flavour factor. This is because 
the Lorentz structure of the two decays are identical. Assuming that 
the decomposition of the $\eta$ which is motivated by experiment, i.e.
$\eta = \frac{1}{\sqrt{2}} (\frac{1}{\sqrt{2}}(u\bar{u}+d\bar{d})+
s\bar{s})$, we have

\beqn \fs = \frac{1}{\sqrt{2}}\fsw = 3.2\pm 0.1\;  \mbox{GeV}  
\hspace{1cm} \fd = \frac{1}{\sqrt{2}}\fdw  = 10.2 \pm 1.5\;   
\mbox{GeV}^{-1} \eeqn

We have also performed a flux--tube ($^3P_0$) model calculation to 
independently
derive the coupling constants. For $\gamma_0 = 0.53$, $\beta = 0.4$ 
GeV we obtain

\beqn \fs =  3.4\; \mbox{GeV} \hspace{1cm} \fd =  10.6\; \mbox{GeV}^
{-1} \eeqn

The agreement (both in sign and magnitude) is clearly impressive, 
underlining the significant agreement of the $^3P_0$ model with 
experiment \cite{geiger94}.

\subsection{$\ggg$}

Here we assume that $100\%$ of the decays of $\rho^+$ are to 
$\pi^+\pi^0$ \cite{pdg96}. We evaluate the amplitude Eq. \ref{rhopipi} 
in the rest frame of $\rho^+$ and connect the amplitude to the width 
of $150.7\pm 0.6$ MeV \cite{pdg96}, according to the usual relation 
(analogous to Eq. \ref{doorwid}) to obtain

\beqn \ggg = 6.02 \pm 0.02 \eeqn

\newpage

\renewcommand{\baselinestretch}{1}

\begin{figure}[t]
\begin{center}
\vspace{-1.5cm}
\caption{\plabel{3ch} Results of the K--matrix analysis.  (a) The events 
($|F_1|^2$) in $\eta\pi$ as compared to experiment \protect\cite{bnletapi};
(b) The phase (of $F_1$) in $\eta\pi$ compared to experiment 
\protect\cite{suhurk}.
The invariant mass $w$ 
is plotted on the horisontal axis in GeV. When the phase is plotted it is 
in radians, with the overall phase {\it ad hoc}. The parameters of the 
simulation are $m_{\rhohat}=1.6$ GeV, $\Gamma_{\rhohat}=168$ MeV 
\protect\cite{bnl97}, $\gamma_1 = 0.31$, $\gamma_2 = 0.52$, 
$\gamma_3 = 1.49$, $m_{b1} = 1.32$ GeV, $m_{b2} = 1.23$ GeV, 
$\gamma_{b1} = 1.53$, $\gamma_{b2} = 2.02$, 
$V_{b1}/V_{\rhohat} = 2.05 e^{2.77 i}$, $V_{b2}/V_{b1} = 0.35 e^{1.6 i}$. 
$V_{\rhohat}$ sets the overall magnitude and phase, which is not shown. 
None of the ratios of production strengths should be regarded as 
physically significant, since the K--matrix formalism allows for the 
introduction of additional parameters in the modelling of the backgrounds, 
which would change the values of these ratios. The plots shown here are 
only weakly dependent
on the $\rho\pi$ parameters $\gamma_{b2}$ and $V_{b2}$.
The parameters have been chosen to fit both the
$\eta\pi$ data \protect\cite{bnletapi} and the preliminary $\rho\pi$ data
 \protect\cite{bnl97}. 
Experiment
has not been able to eliminate the possibility that the low mass peak 
in $\rho\pi$ is 
due to leakage from the $a_1$.
The background
amplitude in $\rho\pi$ is being used as a means of parametrising all forms
 of background into the $\rho\pi$ channel,
including leakage or Deck. 
}
\leavevmode
\vspace{-.7cm}
\hbox{\epsfxsize=3 in}
\epsfbox{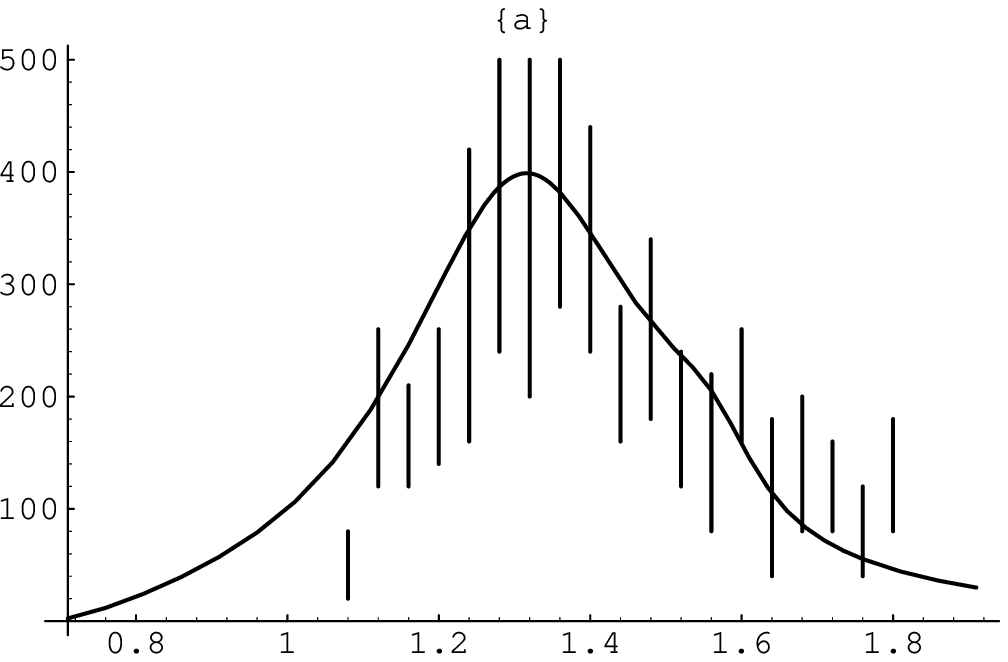}
\vspace{-1.4cm}
\leavevmode
\hbox{\epsfxsize=3 in}
\epsfbox{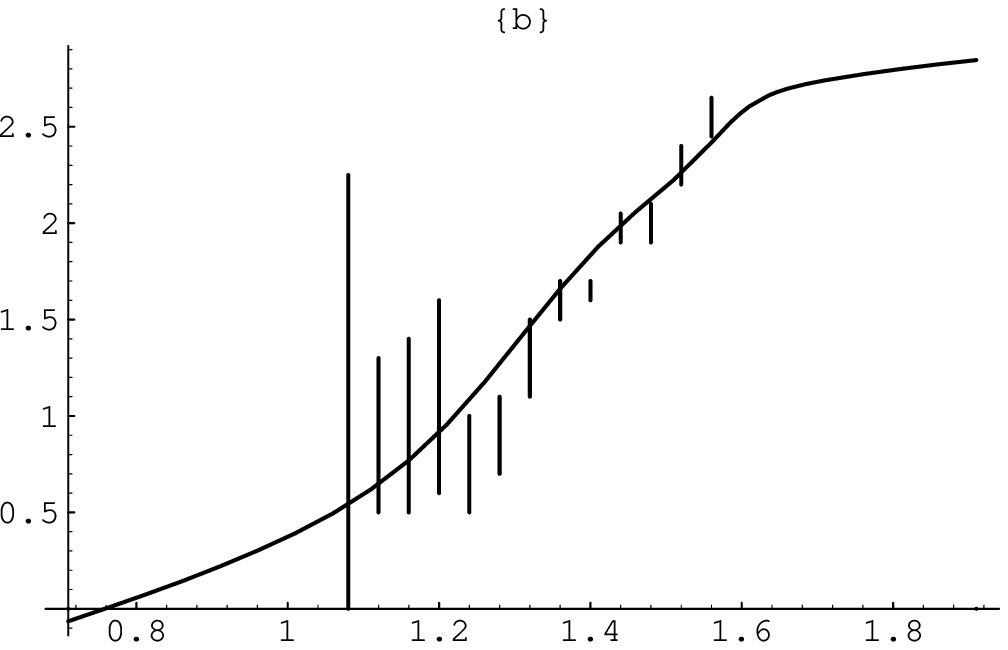}
\end{center}
\end{figure}



\end{document}